
\input harvmac 

\hyphenation{anom-aly anom-alies coun-ter-term coun-ter-terms
dif-feo-mor-phism dif-fer-en-tial super-dif-fer-en-tial dif-fer-en-tials
super-dif-fer-en-tials reparam-etrize param-etrize reparam-etriza-tion}

%
%
\newwrite\tocfile\global\newcount\tocno\global\tocno=1
\ifx\bigans\answ \def\tocline#1{\hbox to 320pt{\hbox to 45pt{}#1}}
\else\def\tocline#1{\line{#1}}\fi
\def\toclead{\leaders\hbox to 1em{\hss.\hss}\hfill}
\def\tnewsec#1#2{\xdef #1{\the\secno}\newsec{#2}
\ifnum\tocno=1\immediate\openout\tocfile=toc.tmp\fi\global\advance\tocno
by1
{\let\the=0\edef\next{\write\tocfile{\medskip\tocline{\secsym\ #2\toclead\the
\count0}\smallskip}}\next}
}
\def\tnewsubsec#1#2{\xdef #1{\the\secno.\the\subsecno}\subsec{#2}
\ifnum\tocno=1\immediate\openout\tocfile=toc.tmp\fi\global\advance\tocno
by1
{\let\the=0\edef\next{\write\tocfile{\tocline{ \ \secsym\subsecsym\
#2\toclead\the\count0}}}\next}
}
\def\tappendix#1#2#3{\xdef #1{#2.}\appendix{#2}{#3}
\ifnum\tocno=1\immediate\openout\tocfile=toc.tmp\fi\global\advance\tocno
by1
{\let\the=0\edef\next{\write\tocfile{\tocline{ \ #2.
#3\toclead\the\count0}}}\next}
}
%
%

%
%
%
%

%
\long\def\optional#1{}
%
%
\let\narrowequiv=\equiv
\def\equiv{\;\narrowequiv\;}
\let\narrowtilde=\tilde
\def\tilde{\widetilde}
\fontdimen16\tensy=2.7pt\fontdimen17\tensy=2.7pt 



%

%
%

%
%
%

%




\def\splitexact#1#2{\mathrel{\mathop{\null{
\lower4pt\hbox{$\rightarrow$}\atop\raise4pt\hbox{$\leftarrow$}}}\limits
^{#1}_{#2}}}

%
%
\def\pa{\partial}

%
%
%
%




%
%

\def\IM{isomorphism}

\def\RS{Riemann surface}

%
%
\def\cmfontflag{cm}
\ifx\fontflag\cmfontflag
\else
 
\fi
\font\blackboard=msym10 \font\blackboards=msym7   
\font\blackboardss=msym5
\newfam\black
\textfont\black=\blackboard
\scriptfont\black=\blackboards
\scriptscriptfont\black=\blackboardss

\let\spec=\blackb
\font\gothic=eufm10 \font\gothics=eufm7
\font\gothicss=eufm5
\newfam\gothi
\textfont\gothi=\gothic
\scriptfont\gothi=\gothics
\scriptscriptfont\gothi=\gothicss

\font\curlyfont=eusm10 \font\curlyfonts=eusm7
\font\curlyfontss=eusm5
\newfam\curly
\textfont\curly=\curlyfont
\scriptfont\curly=\curlyfonts
\scriptscriptfont\curly=\curlyfontss


\global\newcount\figno \global\figno=1
\newwrite\ffile
\def\pfig#1#2{Fig.~\the\figno\pnfig#1{#2}}
\def\pnfig#1#2{\xdef#1{Fig. \the\figno}%
\ifnum\figno=1\immediate\openout\ffile=figs.tmp\fi%
\immediate\write\ffile{\noexpand\item{\noexpand#1\ }#2}%
\global\advance\figno by1}

%
%
\def\tfig#1{Fig.~\the\figno\xdef#1{Fig.~\the\figno}\global\advance\figno by1}








%
%

%


\def\inbar{\,\vrule height1.5ex width.4pt depth0pt}
\def\IB{\relax{\rm I\kern-.18em B}}
\def\IC{\relax\hbox{$\inbar\kern-.3em{\rm C}$}}
\def\ID{\relax{\rm I\kern-.18em D}}
\def\IE{\relax{\rm I\kern-.18em E}}
\def\IF{\relax{\rm I\kern-.18em F}}
\def\IG{\relax\hbox{$\inbar\kern-.3em{\rm G}$}}
\def\IH{\relax{\rm I\kern-.18em H}}
\def\II{\relax{\rm I\kern-.18em I}}
\def\IK{\relax{\rm I\kern-.18em K}}
\def\IL{\relax{\rm I\kern-.18em L}}
\def\IM{\relax{\rm I\kern-.18em M}}
\def\IN{\relax{\rm I\kern-.18em N}}
\def\IO{\relax\hbox{$\inbar\kern-.3em{\rm O}$}}
\def\IP{\relax{\rm I\kern-.18em P}}
\def\IQ{\relax\hbox{$\inbar\kern-.3em{\rm Q}$}}
\def\IR{\relax{\rm I\kern-.18em R}}
\font\cmss=cmss10 \font\cmsss=cmss10 at 10truept
\def\IZ{\relax\ifmmode\mathchoice
{\hbox{\cmss Z\kern-.4em Z}}{\hbox{\cmss Z\kern-.4em Z}}
{\lower.9pt\hbox{\cmsss Z\kern-.36em Z}}
{\lower1.2pt\hbox{\cmsss Z\kern-.36em Z}}\else{\cmss Z\kern-.4em Z}\fi}
\def\IGa{\relax\hbox{${\rm I}\kern-.18em\Gamma$}}
\def\IPi{\relax\hbox{${\rm I}\kern-.18em\Pi$}}
\def\ITh{\relax\hbox{$\inbar\kern-.3em\Theta$}}
\def\IOm{\relax\hbox{$\inbar\kern-3.00pt\Omega$}}

\lref\JDTG{J. Distler, ``2-d quantum gravity, topological field theory,
and the multicritical matrix models,'' Nucl. Phys. {\bf B342} (1990) 523.}%
\lref\EWTG{E. Witten, ``On
the structure of the topological phase of two-dimensional gravity,''
Nucl. Phys. {\bf B340} (1990) 281.}%
\lref\LPWTG{J. Labastida, M. Pernici, and E. Witten, ``Topological
gravity in two dimensions,'' Nucl. Phys. {\bf B310} (1988) 611.}%
\lref\MSTG{D. Montano and J. Sonnenschein, ``Topological strings,''
Nucl. Phys. {\bf B313} (1989) 258; ``The topology of moduli space and quantum
field theory,'' Nucl. Phys. {\bf B324} (1989) 348.}%
\lref\MPTG{R. Myers and V. Periwal, Nucl. Phys. {\bf B333} (1990) 536.}
\lref\BSTG{L. Baulieu and I. Singer, ``Conformally invariant gauge
fixed actions for 2D topological gravity,'' Commun. Math. Phys.
{\bf135} (1991) 253.}%
\lref\VVTG{E. Verlinde and H. Verlinde, ``A solution of two-dimensional
topological gravity,'' Nucl. Phys. {\bf B348} (1991) 457.}%
\lref\EgYa{T. Eguchi and S.-K. Yang, ``N=2 superconformal models as
topological field theories,'' Mod. Phys. Lett. {\bf A5} (1990) 1693.}%
\lref\LSMF{
P. Nelson, ``Lectures on supermanifolds and strings," in
{\sl Particles, strings, and supernovae,} ed. A. Jevicki and C.-I. Tan
(World Scientific, 1989).}
\lref\CIGVO{
P. Nelson, ``Covariant insertions of general vertex operators,"
Phys. Rev. Lett. {\bf62} (1989) 993.}%
\lref\TCCT{J. Distler and P. Nelson, ``Topological couplings and
contact terms in 2d field theory,''
Commun. Math. Phys. {\bf 138} (1991) 273.}%
\lref\semirig{J. Distler and P. Nelson, ``Semirigid supergravity,''
Phys. Rev. Lett. {\bf66} (1991) 1955.}
\lref\DVVCarg{R. Dijkgraaf, H. Verlinde, and E. Verlinde, in {\sl
String theory and Quantum Gravity,} eds. M. Green et al. (World
Scientific, 1990).}
\lref\EWTGlect{E. Witten, ``Two dimensional gravity and intersection
theory on moduli space,'' preprint IASSNS-HEP-90/45.}%
\lref\topsug{S. Govindarajan, P. Nelson and S-J. Rey,
``Semirigid Constructions
of Topological Supergravities,''  UPR0472-T = UFIFT-HEP-91-10
(1991), to appear in Nucl. Phys. B.}
\lref\dil {J. Distler and P. Nelson, ``The Dilaton Equation in
Semirigid String Theory,''  PUPT-1232 = UPR0428T (1991), to appear in
Nucl. Phys. B.}
\lref\sbgpn {S. B. Giddings and P. Nelson,
`` The Geometry of super \RS s,''
Commun. Math. Phys. {\bf 116}, (1988) 607.}
\lref\Polya{A. M. Polyakov, Mod. Phys. Lett. {\bf A2} (1987) 893.}
\lref\KPZ{V. G. Knizhnik, A. M. Polyakov and A. B. Zamoldchikov,
Mod. Phys. Lett. {\bf A3} (1988) 819.}
\lref\DK{J. Distler and H. Kawai, Nucl. Phys. {\bf B321} (1989) 509.}
\lref\LSMS{P. Nelson, ``Lecture on strings and moduli space,'' Phys.
Reports {\bf 149} (1987) 304.}
\lref\BK{E. Brezin and V. A. Kazakov, ``Exactly solvable field
theories of closed strings,'' Phys. Lett. {\bf B236} No. {\bf 2}
(1990) 144.}
\lref\EWKM{E. Witten, ``On the Kontsevich model and other models
of two dimensional gravity,'' IASSNS-HEP-91/24.}
\lref\MATQFT{M. Atiyah, ``Topological quantum field theories,''
IHES Publications Mathematiques {\bf 68} (1988) 175.}
\lref\JPvert{J. Polchinski, ``Vertex operators in the Polyakov path
integral,'' Nucl. Phys. {\bf B289} (1987) 465.}
\lref\JPFBSA{J. Polchinski, ``Factorization of bosonic string
amplitudes,'' Nucl. Phys. {\bf B307} (1988) 61.}
\lref\AGGMV{L. Alvarez-Gaum\'e, C. Gomez, G. Moore and C. Vafa,
Nucl. Phys. {\bf B303}(1988) 455.}
\lref\AGGNSV{L. Alvarez-Gaum\'e, C. Gomez, P. Nelson, G. Sierra
and C. Vafa, Nucl. Phys. {\bf B311} (1988) 333.}
\lref\Segal{G. B. Segal, ``The definition of conformal field theory,''
Differential geometrical methods in theoretical physics (1987) 165.}
\lref\LaNelson{H. S. La and P. Nelson, ``Effective field equations
for fermionic strings,'' Nucl. Phys. {\bf B332} (1990) 83.}
\lref\EWIT{E. Witten, ``Two dimensional gravity and intersection
theory on moduli space," IASSNS-HEP-90/45, to appear in J.
Diff. Geometry.}
\lref\GM{D. J. Gross and A. A. Migdal, ``A nonperturbative treatment of
two dimensional quantum gravity," Phys. Rev. Lett. {\bf 64} (1990) 127.}
\lref\DS{M. Douglas and S.Shenker, ``Strings in less than one dimension,"
Nucl. Phys. {\bf B335} (1990) 635.}
\lref\BDS{T. Banks, M. Douglas and S. Shenker,}
\lref\semigeom{S. Govindarajan, P. Nelson, and E. Wong, ``Semirigid
Geometry,'' UPR-0477T, to appear in Commun. Math. Phys.}
\lref\GNR{S. Govindarajan, P. Nelson, and S. J. Rey, ``Semirigid
construction of topological supergravities," UPR-0472T, to appear in
Nucl. Phys. B.}
\lref\PNpc{Philip Nelson, private communication}
\lref\Manin{Yu. Manin, Gauge field theory and complex geometry,
(Springer-Verlag, 1988).}
\lref\DWMF{R. Dijkgraaf and E. Witten, ``Mean field theory, topological
field theory, and multi-matrix models,'' Nucl. Phys. {\bf B342} (1990)
486.}
\lref\DVVLE{R. Dijkgraaf, Herman Verlinde, and Erik Verlinde, ``Loop
equations and Virasoro constraints in nonperturbative 2-d quantum
gravity,'' Nucl. Phys. {\bf B348} (1991) 435.}
\lref\NSNQLT{N. Seiberg, ``Notes on quantum Liouville theory and
quantum gravtiy,'' 
RU-90-29.}
\ifx\answ\bigans \else\noblackbox\fi

\def\IM{iso\-mor\-phism}

\def\ct{coordinate trans\-formation}

\def\cO#1{{\cal O}_{#1}}
\def\Phg#1{{\hat {\cal P}}_{g,#1}}
\def\Mhg#1{{\hat {\cal M}}_{g,#1}}
\def\Pg#1{ {\cal P}_{g,#1}}
\def\Mg#1{ {\cal M}_{g,#1}}
\def\vt#1{{\tilde v}_#1}
\def\Os{\Omega_\sigma}
\def\Ot{\tilde \Omega}
\def\nuts#1{{\tilde \nu}_#1}
\def\zp{z_P(\cdot)}
\def\sp{{\sigma_P}}
\def\Sp{{\Sigma_P}}
\def\sPsi{|\Psi \rangle}
\def\nut{\tilde \nu}
\def\Sigmah{\hat \Sigma}
\def\>{\rangle}
\def\<{\langle}
\def\tv{{\tilde v}}
\def\mut{{\tilde \mu}}

\Title{\vbox{\hbox{UPR--0491T}}}
{Recursion Relations in Semirigid Topological Gravity}
\centerline{Eugene Wong}\smallskip
\centerline{Physics Department}
\centerline{University of Pennsylvania}
\centerline{Philadelphia, PA 19104 USA}
\bigskip
\bigskip
A field theoretical realization of topological gravity is discussed in the
semirigid geometry context.
In particular, its topological nature is given by the relation
between deRham cohomology and equivariant BRST cohomology and the fact
that all but one of the physical operators are BRST-exact.
The puncture equation and the dilaton equation of pure topological
gravity are reproduced, following reference \dil.
\vskip1cm
\Date{11/91}\noblackbox
\newsec{Introduction}

In four dimensions, field theories of quantum gravity break down due to
non-renormalizability, signaling the need for a more fundamental
theory.
This is however not the case in two dimensions.  It is adequate to
describe 2-d quantum gravity within the context of a field theory
\Polya \KPZ \DK.
Moreover, if 2-d gravity is exactly solvable then
it gives us a handle to search
for qualitative features
that may persist in 4-d \NSNQLT.
However, a renormalizable theory does not imply that it is solvable, and
here the difficulty lies in the Liouville mode of 2-d quantum gravity.
The Liouville
mode decouples in a modified version of 2-d quantum gravity
called topological gravity, much
like string theory in the critical dimension.

In the path integral context, quantizing gravity amounts to integrating
over all different metrics $g_{\mu \nu}(x)$.
This integration possesses two subtleties,
namely what we mean by ``all'' and what we mean by
``different''.  Certainly, we must at least include all of the moduli
space: the space of all metrics
modulo coordinate transformations and conformal transformations
when scaling is also a true symmetry.
This moduli space is equivalent to the space of all conformal structures
modulo coordinate transformations (see eg. \LSMS).
There exists yet another definition of this moduli space, namely by
the universal family of stable algebraic curves \LSMS.
In rough correspondence with these three constructions of moduli space,
three main classes of modified 2-d gravity have emerged, each simpler
to deal with than pure Liouville gravity.  Corresponding to the sum
over all metrics, we have discretizations of spacetime at some
critical point, yielding various matrix models \BK \GM \DS; to the sum over
conformal structures we associate critical field theoretic
realizations of topological gravity (see e.g. \LPWTG \VVTG \semirig).
Finally, corresponding to the space of stable curves, one is led to consider
the topological invariants of this moduli space \EWTG.

These theories are all related to the 2-d quantum gravity
Polyakov first wrote down but the relations
between the elementary variables among the
different approaches are not so clear.
Moreover, there is an added complication of
coupling matter to gravity which may naturally
appear in the above approaches.
However, the end results of correlation functions
of observables in all cases seem to reproduce
the same physics, in particular, when pure topological
gravity is concerned \EWTG\VVTG\DWMF.
In this paper we will concentrate on
pure topological gravity.

Let us briefly recall each of the three modifications to pure gravity
mentioned above, starting with a description of the different field theoretical
realizations of topological gravity.
By now, there are numerous topological quantum field theories of
2-d gravity obtained in different ways. \foot{A partial list
includes \LPWTG \MSTG \MPTG \BSTG \VVTG \JDTG.}
What these theories lack in summary is
an underlying geometrical principle to define and obtain
a topological field theory of gravity and compute correlation
functions.

In this paper, we follow the
approach taken by Distler and Nelson \semirig\dil\ which is best
described in the following interpretation.
Field theories live on spacetime manifolds.
Integrating over inequivalent spacetime manifolds then gives
quantum gravity.
In Polyakov's theory, conformal symmetry is not exact.
Therefore, quantum gravity is given by integrating over all inequivalent
complex spacetime manifolds and also the Liouville mode.
To obtain topological gravity from 2-d quantum gravity,
Distler and Nelson proposed a new spacetime geometry which again possesses
conformal symmetry.
However, besides having complex structures on a 2-d manifold, they
suggested that this
manifold also possesses a ``semirigid structure'' \semirig \semigeom.

One constructs semirigid geometry starting with a geometry
associated to local
$N=2$ supersymmetric 2-d gravity.  One then
constrains and twists the two supersymmetry transformations
to leave only the semirigid symmetry transformations \semirig.
Since supersymmetry is a spacetime symmetry, so is the semirigid symmetry.
The moduli space now becomes the space of all {\it semirigid} complex
manifolds modulo
isomorphisms due to \ct s.
This semirigid moduli space is
well understood \semirig \dil \semigeom.
On these semirigid manifolds, a
topological field theory of gravity can be defined.
{}From this point of view, we can imagine
a phase transition at the Planck scale to the new semirigid
symmetry as opposed to zero vacuum expectation of the metric \EWTG.
The stress energy tensor,  with its topological character, and the BRST
charge come out naturally in superfields.
One can then apply the operator formalism
\AGGMV\AGGNSV\ and construct the correlation
functions of non-trivial observables given by \VVTG.

There are two approaches other than field theoretical mentioned above
which also
simplify Polyakov's quantum gravity and one of them is the matrix
models.  We will show
in this paper that in two respects, correlation functions obtained in
the semirigid
formulation reproduce those of a particular matrix model.
Matrix models provide an alternate way to sum all the
metrics on spacetime lattices.  In these discrete models, the volume
of diffeomorphisms gets replaced by a finite factor much as in
lattice gauge theories, so one needs not fix a gauge.
One then takes the continuum limit and
obtains theories of quantum gravity (and sometimes coupled to matter).
To establish the link between topological gravity and matrix
models, we recall some results of
the relevant matrix model.
The one matrix model with an $N \times N$ hermitian matrix corresponds to a
dynamically triangulated random surface. This matrix model is
generalized \BK\ to correspond also to surfaces generated by squares,
pentagons, etc.  Such a one matrix model exhibits multicritical
behavior indexed by the integer $k$.
For example, in the large $N$ and at the $k = 2$
critical point double scaling limit, it gives the same scaling relations
as the Liouville theory of pure quantum gravity \BK.
At the $k = 1$ critical point in the double scaling limit, the one
matrix model reproduces
topological gravity of various other formulations  \EWTG\VVTG\JDTG.
In principle, this $1 / N$ expansion in the double scaling limit of
the matrix model provides a
non-perturbative definition of 2-d quantum gravity.
However, the results can of course be expanded in the string coupling to get an
expansion in the number of handles $g$ of individual surfaces.
Correlation functions in this perturbative expansion obey recursion
relations \DVVLE.  In particular, at the $k=1$ critical point, one has
the puncture
$\cO0$ and dilaton $\cO1$
equations,
\eqn\ePE{\< \cO0 \cO{n_1} \ldots \cO{n_N}\>_g =
\sum_{i=1}^N n_i \< \cO{n_i -1} \prod_{j \not= i}^N  \cO{n_j}\>_g \quad
{\rm and}}
\eqn\eDE{ \< \cO1 \cO{n_1} \ldots \cO{n_N}\>_g =
(2g-2+N) \< \cO{n_1} \ldots \cO{n_N}\>_g.}

In \eDE, we recognize the prefactor as a topological
invariant, the Euler number of an $N$ punctured Riemann surface.
In this paper, as a sequel to \dil, we use the field theoretical
method in the operator formalism \AGGMV\AGGNSV\ to show
how the boundary of moduli space of $N=0$ gravity can
contribute to give topological coupling \ePE\ and how
the bulk of the moduli space contributes the factor of
$2g-2$ in \eDE.  The contact bit of \eDE\ was studied in \dil.
Hence we will show that two recursion relations in the one matrix
model at the $k=1$ critical point are reproduced in the semirigid
topological field theory.
Other recursion relations involve \RS s with different genera \DVVLE.
To recover non-perturbative physics, one turns these
recursion relations into differential equations and
shows that their solutions will contain the same non-perturbative
information as in the one matrix model \DVVCarg.

Finally, the third approach to simplifying 2-d gravity mentioned above
was intersection theory.  In such theory, one computes the topologically
invariant intersection numbers of certain subspaces of the space of
stable algebraic curves.  There exist
established intersection theories to
compute these topological invariants \EWTG.  In \DWMF,
the puncture equation \ePE\ is derived for arbitrary genus
from the intersection theory following Deligne.
More recently, Witten has shown \EWKM\ indirectly that the other recursion
relations are true as well by establishing a link between Kontsevich's
formulation of intersection theory and the one matrix model.

These intersection numbers obey the axioms
of a topological quantum field theory \MATQFT\EWIT.
Furthermore, they are independent
of any specific field theoretical implementation.
This generality however is also a reason why the intersection theories are more
abstract and hence more
difficult to compute than in a specific field theoretic
realization of topological gravity.  Here, we trade
this abstractness with a field theoretic calculation that is simple
and is valid for all genera.

The paper is organized as follows.  In section 2, we review the
semirigid geometry and describe its moduli space. In section 3, we
introduce an operator formalism which is used to construct correlation
functions.  The relation between BRST cohomology on the Hilbert space
and the deRham cohomology on the moduli space is discussed.  We end
the section
by giving the observables in topological gravity first obtained by
E. Verlinde and H. Verlinde.  In section 4, the
dilaton equation \eDE\ is computed ignoring contact terms and in section 5,
the puncture equation \ePE\ is derived.

\newsec{Semirigid geometry}
In this section, we will review semirigid
geometry \semirig\dil\GNR\semigeom.
In particular, we will show how to impose a semirigid
structure on an $N=2$ super Riemann surface,
obtain the stress energy tensor and
the BRST charge.
The semirigid moduli space is then described
and the sewing prescription is given.  The sewing
prescription is needed for the contact term contributions
to \ePE\ and \eDE.

\subsec{Semirigid gravity from N=2 superconformal geometry}

An $N=2$ super Riemann surface is patched from pieces of ${\spec C}^{1|2}$ with
coordinate $(z, \theta, \xi)$.  The transition function
on an overlap is given by the superconformal coordinate
transformation \semirig,
\eqn\entct{\eqalign{ z' &= f + \theta t \psi + \xi s \tau
+ \theta \xi \pa (\tau \psi) \cr
\theta' &= \tau + \theta t + \theta \xi \pa \tau ,\quad
\xi' = \psi + \xi s - \theta \xi \pa \psi ,}}
where $f,t,s$ ($\tau, \psi$) are even (odd) functions of $z$ and
restricted to transformations with
$\pa f = ts - \tau \pa \psi - \psi \pa \tau$.

The $N=2$ super Riemann geometry only dictates that $\theta \xi$
be of spin one, like the coordinate $z$.  The spin of $\theta$ is
not defined a priori.
To obtain a spin zero $\theta$ geometrically, we restrict to
transformations with $\tau = 0$ and $t=1$
so that $\theta$ does not change under coordinate
transformation.  We are left with
\eqn\esrct{\eqalign  {z' &= f + \theta \psi, \quad
  \theta' =  \theta   \cr
   \xi' &= \psi + \xi \pa f - \theta \xi \pa \psi. } }
On a super manifold with patching functions of this form,
$ D_\theta = {\pa \over {\pa \theta}} + \xi{\pa \over {\pa z}}$
becomes a global vector field and
$\{{\tilde D}_\xi=
 {\pa \over {\pa \xi}} + \theta{\pa \over {\pa z}} \}$
span a line
bundle ${\cal D_-}$.  We can thus mod out the flow of ${\cal D_-}$ and
obtain semirigid
coordinate transformations in terms of $(z, \theta)$, $\theta$ being
spin zero.

An infinitesimal $N=2$ coordinate transformation is generated by
$$V_{{\narrowtilde v}^z} = {\tilde v}^z \pa_z
+ {1\over 2}(D {\tilde v}^z ){\tilde D}
+ {1\over 2}({\tilde D}  {\tilde v}^z )D$$
where ${\tilde v}^z = {\tilde v}^z(z, \theta,\xi)$
is an even tensor field.
Since $D$ does not transform on a semirigid surface, we impose that
$\tv^z$ satisfies
${\tilde D} {\tilde v}^z =0$, hence
\eqn\evtz{{\tilde v}^z=v_0^z + \theta \omega^\xi
+ \theta \xi {v_0^z}' .}
Substituting \evtz\ back into $V_{{\narrowtilde v}^z}$, it generates
the infinitesimal
version of semirigid coordinate transformation \esrct.

To define a field theory on the semirigid manifold, we begin with the
(twisted) $N=2$ superconformal
ghosts and their stress energy tensor,
\eqn\eghosts{C^z = c^z + \theta \gamma^\xi
+ \xi {\check \gamma}^\theta + \theta \xi {\check c} ,\quad
B_z = {\check b}_z + \theta {\check \beta}_{\theta z}
- \xi \beta_{\xi z} + \theta \xi( b_{zz} + \pa_z {\check b}_z),}
and
\eqn\ecSten{ T_z = J_z + \theta {\tilde G}_{\theta z} - \xi G_{\xi z}
+ \theta \xi ( (T_B)_{zz} + \pa_z J_z) .}
%
Keep in mind that
$\theta$ and $\xi$ in \eghosts\ and \ecSten\ are spin zero and one
respectively.
We have the desired supersymmetric partners of the same spin, namely
$(b_{zz},\beta_{\xi z})$ and $(c^z,\gamma^\xi)$, but we also have the
unwanted fields $({\check b}_{z},{\check \beta}_{\theta z})$ and
$({\check c},{\check \gamma}^\theta)$.
Therefore we will
constrain the theory and eliminate half of the degrees of freedom.
Since ${\check \gamma}^\theta$ is a scalar field, it make sense to
constrain it to be a constant.
To be consistent, this constraint must follow from a superfield
constraint.  This leads us to impose \semirig
\eqn\econst{{\tilde D}_\xi C^z = q}
where $q$ is a constant.  This superfield constraint breaks the full
$N=2$ symmetry group down to the
subgroup \esrct\ since the latter preserves $D_\theta$
and the lhs of \econst\ $({\tilde D}_\xi C^{\xi \theta})$
transforms dually to $D_\theta$.
Equation \econst\ implies that ${\check \gamma}$ is the constant $q$
and ${\check c}$ is not an
independent field, ${\check c} = \pa c$.
The definition of the components of $B_z$ in \eghosts\ is twisted
so that $({\check b},{\check \beta})$ are conjugate to $({\check
c}.{\check \gamma})$ which are eliminated by \econst.
Moreover, the zero mode insertions which we will use in section 3 to
define correlation functions are of the form
$$ \oint[dz d\theta d\xi] B_z \tv^z=- \oint dz (\beta_{\xi z}
\omega^\xi + b_{zz} v_0^z)$$
since $\tv^z$ that generates infinitesimal semirigid
coordinate transformation is given by \evtz.
Hence if we consider inserting only operators independent of
$({\check b}_{z},{\check \beta}_{\theta z})$ and
$({\check c},{\check \gamma}^\theta)$, then these fields will decouple
altogether from the theory.

To generate coordinate transformation \esrct,
only part of the stress energy tensor \ecSten\ is needed.
The definition of the bosonic energy tensor $T_B$ in \ecSten\ is twisted
so that the unbroken generators of \esrct\ are modes of
$G_{\xi z}$ and $T_B$.
To see that we again use $\tv^z$ of \evtz\ which generates
infinitesimal semirigid coordinate transformation and obtain the
corresponding generators of the stress energy tensor \ecSten\ given by
$$ -\oint[dzd\theta d\xi]~T_z\tv^z =
\oint dz (T_B v_0^z + G_{\xi z}\omega^\xi).$$
This twisting as in \EgYa\
will lead to an anomaly free theory, as we will see.

The full $N=2$ stress energy tensor is given by \semirig
\eqn\eSten{ T_z = \pa (CB) - {1 \over 2} (DB \tilde D C + \tilde D BDC) }
and the BRST charge
\eqn\eBRST{ Q = - {1 \over 2} \oint [dz d \theta d \xi] C^z T_z.}
Imposing the constraint \econst, we obtain in components
the unbroken generators by substituting \eghosts\ and \ecSten\ into
the constrained
\eSten\ and \eBRST,
\eqn\eTbc{ T_B = -2b \pa c - (\pa b) c - 2 \beta \pa \gamma
- (\pa \beta) \gamma,}
\eqn\eGc{ G_{\xi z} = -2 \beta \pa c - (\pa \beta) c + {q \over 2} b,}
and
\eqn\eBRSTc{\eqalign { Q_T &= - {1 \over 2} \oint dz (- c^z T_{zz}
+ \gamma^\xi
G_{\xi z} + q {\tilde G}_{\theta z}) \cr
  &= \oint dz \lbrack - c b \pa c + \beta \gamma \pa c
- \beta c \pa \gamma
- {q \over 2} b \gamma \rbrack.} }
Since $(\beta, \gamma)$ have the same spin as $(b,c)$
but the opposite statistics, the central charges from
them cancel.  Moreover, note that in \eGc\
and \eBRSTc, $G$ and $Q_T$ differ from their $N=1$ counterparts.
Using the $N=1$ superfield expressions and replacing the
spin one half $\theta$ by spin zero $\theta$
give incorrect answers.

Next, we expand in modes $L_n = \oint T_B z^{n+1} dz$ and
$G_n= \oint G_{\xi z} z^{n+1} dz$ and similarly for the ghosts $(b,c)$
and $(\beta, \gamma)$.  Imposing the usual commutation
relations $[b_m,c_n] = \delta_{m+n,0}$ and
$[\gamma_m,\beta_n]=\delta_{m+n,0}$ where $[~,~]$ is the graded bracket,
we obtain the following algebra
\eqn\ealgebra{\eqalign
{\lbrack L_m , L_n \rbrack &= ( m - n) L_{m+n}, \quad
\lbrack L_m , G_n \rbrack = ( m - n) G_{m+n} , \quad
\lbrack G_m , G_n \rbrack  = 0,   \cr
\lbrack G_m , b_n \rbrack &= ( m - n) \beta_{m+n} , \quad
\lbrack G_m , c_n \rbrack = {q \over 2} \delta_{m+n,0},   \cr
\lbrack G_m , \gamma_n \rbrack &= - ( 2m + n) c_{m+n}, \quad
\lbrack G_m , \beta_n \rbrack = 0 ,  \cr
\lbrack L_m , b_n \rbrack &=  ( m - n) b_{m+n} , \quad
\lbrack L_m , c_n \rbrack  =  - ( 2m + n) c_{m+n},  \cr
\lbrack L_m , \gamma_n \rbrack &=  - ( 2m + n) \gamma_{m+n} , \quad
\lbrack L_m , \beta_n \rbrack =  ( m - n) \beta_{m+n},  \cr
\lbrack Q_T , b_n \rbrack &=  L_n , \quad
\lbrack Q_T , c_n \rbrack  =
\sum_m (m-n+1) c_m c_{n-m} -{q\over 2}\gamma_n, \cr
\lbrack Q_T , \beta_n \rbrack &=  - G_n,
\quad {\rm and} \quad
\lbrack Q_T , \gamma_n \rbrack =  \sum_m (2m-n) c_m \gamma_{n-m}.    }}
We set $q=-2$ as in \dil\ and \VVTG.
{}From the commutator of the BRST charge with the mode $b_n$,
we obtain $\lbrack Q_T, b(z) \rbrack  = T_B(z)$.
The stress energy tensor being BRST-exact is the
signature of a topological theory, implying
that the metric dependence of the action
decouples \EWTG.

\subsec{Semirigid moduli space}

We will now discuss the semirigid moduli space following \semirig
\dil\GNR\semigeom.
To understand the semirigid moduli space, we introduce
a family of augmented surfaces.   To build an augmented surface, we
start with
an ordinary \RS\ given by patching maps $z'=f(z)$.  We then
introduce a new global spinless anticommuting coordinate $\theta$ and promote
all patching functions to superfields in $\theta$.  Thus, an
augmented surface is obtained with patching maps
\eqn\eact{z'=f(z, \theta ) \equiv f(z) + \theta \phi (z);
\quad \theta'=\theta.}
This is the same as \esrct\ after modding out ${\cal D}_-$ when
we identify $\phi(z)$ with $\psi(z)$ of \esrct.  Hence, surfaces
patched together
by the augmented maps have a one to one correspondence with the semirigid
surfaces
since given either patching function, we can recover the other \GNR\semigeom.

Suppose now we have a {\it family} of \RS s parametrized by ${\vec
m}$, that is
$z' =f(z;{\vec m})$ are the patching maps.  To obtain a family of
semirigid surfaces, we
merely need a family of augmented \RS s.
We let the family of augmented \RS s have patching functions $\theta ' =
\theta$ and
\eqn\eamod{z'=f(z;{\vec m}+ \theta {\vec \zeta})
=f(z;{\vec m}) + \theta\zeta^i h_i(z;{\vec m})}
where $h_i(z;{\vec m}) =  \pa_{m^i} f(z;{\vec m})$.
The original moduli ${\vec m}$ are augmented by
$\theta{\vec \zeta}$ giving an equal number of odd moduli ${\vec \zeta}$.
One can now show using \eamod\ that another parametrization $({\vec m}')$
of the {\it same} family of surfaces induces $({\vec m}'),{\vec \zeta}')$
which are related to the original ones by a split coordinate
transformation.
That is, ${\vec m}' = {\vec m}'({\vec m})$ and
${\zeta '}^i  = (\pa_{m^j} m'^i) \zeta^j$ \dil\semigeom.
We will use this property later on.

Consider the moduli space of genus $g$ semirigid surfaces
with one puncture at $P$,
$${\hat {\cal M}}_{g,1} \equiv
{[{\rm all~semirigid~complex~manifolds~with~puncture}] \over
[{\rm isomorphisms~preserving~puncture}]}.$$
It has a natural projection to the unpunctured moduli space
$${\hat {\cal M}}_{g,0} =
{[{\rm all~semirigid~complex~manifolds}] \over [{\rm isomorphisms}]}$$
simply by forgetting $P$
More generally, we can have a moduli space of genus $g$
semirigid surfaces with $N$ punctures $\Mhg{N}$.
The integration density on $\Mhg0$ is interpreted to be
the integrand in the path integral without source terms.
Hence, the integral over $\Mhg0$ of the volume density
gives the partition function.
The one point correlation function is then the integral of a certain
integration density defined on $\Mhg1$ as we will recall below.
One of the consequences of having a split
coordinate transformation
on semirigid moduli space $\Mhg{N}$ is that there also exists a
natural projection $\Pi : \Mhg{N} \rightarrow \Mg{N}$ to the
ordinary moduli space of \RS s with $N$ punctures \dil.
This means that if we have a measure
on $\Mhg{N}$, then we can without further obstruction integrate along
the fibers $\Pi^{-1}$ (ie. integrate out all the odd moduli
${\vec \zeta}~$) leaving a measure on the ordinary moduli space \dil.

A bundle $\Phg1$ with base space $\Mhg1$ can be
constructed \AGGMV\LaNelson, where the
fiber over each point $(\Sigmah ,P) \in \Mhg1$
consists of the germs of coordinate systems
$z_P(\cdot)$ on $\Sigmah$ defined near $P$ with their origins at $P$.
(We always take $\theta_P(\cdot) = \theta(\cdot)$, the global odd
coordinate on $\Sigmah$.)
Thus, we have ${\hat \pi} : \Phg1 \rightarrow \Mhg1$.
The analog of the Virasoro action
on ordinary ${\cal P}$ \AGGMV\
is defined by the infinitesimal form
of coordinate transformation \eact\ on $z_P(\cdot)$. If we let
$f(z)= z-\epsilon z^{n+1}$ and $\phi(z) = \alpha z^{m+1}$ in \eact, where
$\epsilon$  and $\alpha$ are commuting and anticommuting infinitesimal
parameters respectively, then the corresponding generators are defined
as $l_n =-z^{n+1} \pa_z$ and $g_m=-\theta z^{m+1} \pa_z$ so that
$z' = (1 + \epsilon l_n + \alpha g_n)z$.
$l_n$ and $g_m$ are the generators of the augmented version of the
Virasoro group.
In the following, we denote a semirigid surface $\Sigmah$
with a unit disk $|z_P(\cdot)|\le 1$ centered at $P$
removed by $(\Sigmah, z_P(\cdot))$, and similarly for a \RS\ with a unit disk
removed $(\Sigma,z_P(\cdot))$.

We will review some facts about a vector field
$\vt{i} \in T\Pg1$ for ordinary geometry $\pi: \Pg1 \rightarrow \Mg0$
\AGGMV.

Given $\Sigma$, we can
deform it to a neighboring $\Sigma'$ by
some $\vt{i} \in T\Pg1$. We will classify into three categories the
action of a Virasoro generator
$v=-\epsilon \sigma^{n+1} \pa_\sigma$, where $\sigma=\zp$ is some local
coordinate.  We denote the Virasoro action
on $\Pg1$ by
$i_\sigma (v) = \tv$, a tangent to $\Pg1$ at $\sigma$.
To construct $\Sigma'$, we begin with $(\Sigma,\sigma)$.
We then identify points on the boundary of
$(\Sigma,\sigma)$ with those of a unit disk $D$
via a composition of $\sigma$ with the map
$1+ v$, yielding $\sigma' = \sigma - \epsilon \sigma^{n+1}$, a
``Schiffer variation'' of $(\Sigma,\sigma)$.
If $v$ extends analytically to $\Sigma \backslash P$
(eg. $n \le 1-3g$), then $\Sigma '$
is identical to $\Sigma$ because the variation can be undone by a
coordinate transformation generated by $v$ on the rest of the surface
$\Sigma$.  Thus, $i_\sigma (v) = 0$.  If $v$ extends holomorphically to $D$
and vanishes at $\sigma=0$ ($n \ge 0$),
then we merely have a coordinate transformation
on $\sigma$.  $i_\sigma (v)$ is then vertical along the fiber in
$\Pg1$ and $\pi_*$
kills it.
The Weierstrass gap theorem \AGGMV\ states that on an unpunctured
Riemann surface
with $g > 1$,
every meromorphic vector
field $v$ on the disk $D$ can be written as the sum of a holomorphic
vector on $D$ and a vector field that extends to the rest of $\Sigma$,
except for a $(3g-3)$ dimensional subspace.
The $3g-3$ dimensional subspace of vector fields have simultaneously
a pole in $D$ and in the rest of $\Sigma$ (eg. $-2 \ge n \ge 2-3g$).
This $3g-3$ dimensional vector space when projected down to $\Mg0$ by
$\pi_*$ yields the full $T\Mg0$,
hence the ordinary moduli space $\Mg0$ is $3g-3$ dimensional.  By the
augmented construction of semirigid moduli space, we similarly see
that $\Mhg0$ is
$3g-3|3g-3$ dimensional \semirig.
Finally, $\tv = \epsilon \pa_\sigma \in T\Pg1$ gives a vector field that moves
the puncture $P$.

Consider a family of once-punctured semirigid surfaces
with a family of semirigid coordinates $\sigma$ centered at the puncture $P$,
constructed by augmenting a similar family of ordinary \RS s as in
\eamod.  Thus, as shown in figure 1, we have
$\sigma = f(z;{\vec m}) + \theta \zeta^i h_i(z;{\vec m})$
where $h_i={\pa_{m^i}}f$ and $({\vec m},{\vec \zeta}~)$ the coordinates
for the $3g-2|3g-2$ dimensional $\Mhg1$.
We will see that a puncture in semirigid geometry has one even and one
odd modulus associated with it and
hence the dimension of $\Mhg1$ is $1|1$ bigger than $\Mhg0$.
Keep in mind that $\sigma$ has to be holomorphic in $z$ and $\theta$
but not necessary in ${\vec m}$.
Then for $k = 1,\ldots,3g-2$,
\eqn\epushfe{\eqalign{ \sigma_* (\pa_{m^k}) &=
{\pa \sigma \over \pa m^k} {\pa \over \pa \sigma} +
{\pa {\bar \sigma} \over \pa m^k} {\pa \over \pa {\bar\sigma}} \cr
&= (\pa_{m^k} f +\theta \zeta^i \pa_{m^k} h_i) \pa_\sigma +
(\pa_{m^k} {\bar f} +{\bar \theta}{\bar \zeta^i} \pa_{m^k} {\bar h_i})
\pa_{\bar \sigma} \equiv \vt{k}  }}
is the push forward of an {\it even} vector $\pa_{m^k} \in T\Mhg1$ to $\vt{k}
\in T\Phg1$ and
\eqn\epushfo{\sigma_*(\pa_{\zeta^k}) = - \theta h_k \pa_\sigma \equiv
\nuts{k}}
is the push forward of an {\it odd} vector $\pa_{\zeta^k} \in T\Mhg1$ to
$\nuts{k} \in T\Phg1$.
Note that in
$\tv_k = \vt{k}^z \pa_z + {\bar \vt{k}}^{\bar z}\pa_{\bar z}$ of \epushfe,
${\bar \vt{k}}^{\bar z}$ is not necessary the complex conjugate of
$\vt{k}^z$; also, $\nuts{k}$ of \epushfo\ is proportional to $\theta$,
a special property of the augmented coordinates.
These vectors $(\vt{k}, \nuts{k})$ when projected down by ${\hat \pi}_*$
span the $3g-2|3g-2$ dimensional holomorphic tangent space $T\Mhg1$
analogous to the
vectors $\pi_*\vt{k}$ with $\vt{k} \in T\Pg1$ in the above discussion
of ordinary geometry.

We now give the prescription for sewing two semirigid surfaces that
is compatible with the compactification of moduli space by stable curves.
In superspace, a ``point" $P$ is defined by the vanishing of some functions.
In particular, if we have some even function $f(z,\theta)$ on a
semirigid surface, then $P$ can
be defined by where $f = \theta =0$ \semirig.  Note that with any invertible
function $g$, $g \cdot f$ defines the same $P$.  A divisor in
the semirigid superspace is thus given by a semirigid coordinate
$z_P(\cdot) = z - z_0 - \theta \theta_0$  centered at $P$. Higher
Taylor coefficients in $z$ do not matter since they can be introduced or
removed freely by $g$.
$(z_0,\theta_0)$ are the $1|1$ parameters associated to the position of $P$.
Consider a semirigid surface $\Sigmah$ with genus $g$
without punctures degenerating into two pieces $\Sigmah_L$ and
$\Sigmah_R$ of genera $g_L$ and $g_R$.  Let $P_L$ and $P_R$ be the
double points at the node on $\Sigmah_L$ and $\Sigmah_R$.  Counting the
number of moduli of $\Sigmah$ $(3g-3|3g-3)$
and the sum of that of $(\Sigmah_L,P_L)$ $(3g_L-2|3g_L-2)$
and $(\Sigmah_R,P_R)$ $(3g_R-2|3g_R-2)$,
we have an excess of $1|1$ dimension in $\Sigmah$'s moduli space
over that of $\Sigmah_L$ and
$\Sigmah_R$.  Hence the plumbing fixture joining the two semirigid surfaces
$\Sigmah_L$ and $\Sigmah_R$ must depend on a $1|1$ sewing moduli.
The augmented sewing prescription tells us to join the two surfaces by
relating the
coordinates $z_L|\theta_L$ and $z_R|\theta_R$ centered at the sewing
points $P_L$ and $P_R$ in the following way \semirig\dil,
\eqn\eplumb{z_L = {q+\theta_R \delta \over z_R},
\quad \theta_L=\theta_R.}
$(q,\delta)$ are the $1|1$ sewing moduli
because changing $(q,\delta)$ alter
the resultant surface $\Sigmah$.
Since \eplumb\ is of the form \esrct, we get a semirigid surface after
sewing.

We introduce the plumbing fixture because we want to complete the
moduli space to a compact space.  For this, we need a precise
description of how semirigid surfaces may degenerate.
The stable curve compactification of the moduli space
requires that $\Sigmah_L$ and $\Sigmah_R$ be independent of $(q,\delta)$
and that this dependence come solely from sewing in the limit $q
\rightarrow 0$.  Away from $q \rightarrow 0$, it does not matter
whether $\Sigmah_{L,R}$ depend on $q$.

Another rule from the stable curve compactification is that no two nodes of
$\Sigmah$ can collide and neither can two punctures.
Two colliding points $P_1$ and $P_2$ on $\Sigmah$ is replaced by a
conformally equivalent degenerating surface.
The conformally equivalent surface consists of the same surface
$\Sigmah$ with one puncture $P_L$ at the would be colliding point
identified with the puncture $P_R$ of
a fixed three punctured sphere ${\bf P}^1_{(P_R,P_1,P_2)}$.
Here, we see that we can apply the sewing
prescription with $\Sigmah_L = \Sigmah_{P_L}$ and  $\Sigmah_R =
{\bf P}^1_{(P_R,P_1,P_2)}$ via the plumbing fixture \eplumb.  The
three punctured sphere is rigid and
has no moduli associated with it \semirig.  The moduli from the
plumbing fixture
are associated to the distance between $P_1$ and $P_2$.

The most general coordinates near the three punctures $(P_R,P_1,P_2)$
centered at
$(\infty, E, {\tilde E})$ on the sphere can be given by \TCCT\dil
\eqn\esphere{ ({\bf P}^1;~ z^{-1},~ (z-E)+a_1(z-E)^2 +a_2(z-E)^3+\ldots,~
   (z-{\tilde E})+ {\tilde a}_1(z-{\tilde E})^2 +\ldots).}
Let the coordinate centered at $P_L$ be $(\sigma,\theta)$.
Then by \eplumb\ with $z_L=\sigma$ and $z_R=z^{-1}$, we can express $z$
in terms of $\sigma$, $z=q^{-1}\sigma(1-q^{-1} \theta \delta)$.
Substituting this back into \esphere, we obtain the most general
coordinates centered at the colliding $P_1$ and $P_2$ using the local
coordinates $(\sigma,\theta)$.  They are
\eqn\egsc{z_{P_1}(\cdot)= q^{-1}\sigma (1- q^{-1}\theta \delta)-E
+   a_1 [q^{-1}\sigma (1- q^{-1}\theta \delta)-E]^2 +\ldots,}
and similarly for $z_{P_2}$ with $a_i \rightarrow {\tilde a}_i$ and
$E \rightarrow {\tilde E}$.
These two coordinate slices will be used in calculating the contact
terms in section 5.

\newsec{Operator Formalism}

\def\cO#1{{\cal O}_{#1}}
\def\Phg#1{{\hat {\cal P}}_{g,#1}}
\def\Mhg#1{{\hat {\cal M}}_{g,#1}}
\def\Mg#1{ {\cal M}_{g,#1}}
\def\vt#1{{\tilde v}_#1}
\def\Os{\Omega_\sigma}
\def\Ot{\tilde \Omega}
\def\nuts#1{{\tilde \nu}_#1}
\def\zp{z_P(\cdot)}
\def\sp{{\sigma_P}}
\def\Sp{{\Sigma_P}}
\def\sPsi{|\Psi \rangle}
\def\nut{\tilde \nu}
\def\Sigmah{\hat \Sigma}
\def\>{\rangle}
\def\<{\langle}
Following G. Segal's construction \Segal, a conformal field theory associates a
state
in the Hilbert space $|\Sigmah,z \>$ to a one-punctured semirigid Riemann
surface with local coordinate $z$ denoted by $(\Sigmah, z)$.
Recall that $z$ is a coordinate we put on $\Sigmah$ at the
point where $z=0$ and a unit disk $|z| \le 1$ has been removed from
the semirigid surface.
Since a state in the Hilbert space associated to
$(\Sigmah , z)$ depends on $z$,
the Virasoro
action previously defined on $\hat {\cal P}$ also acts on the
Hilbert space.
Following \LaNelson, we define
using mode expansions $L_n$, $G_m$ of $T_B$, $G_{\xi z}$ and their
conjugates
\eqn\estatem{\langle \Sigmah, z-\epsilon z^{n+1} -\alpha \theta z^{m+1}|
             =\langle \Sigmah, z|~ [(1+ \epsilon L_n + \alpha G_m)
      (1+ {\bar \epsilon} {\bar L}_n + {\bar \alpha} {\bar G}_m) + \ldots],}
where the ellipsis refer to terms of order $\epsilon^2$, etc..

We will now discuss how to obtain a measure on
$\Mhg1$ from $\Phg1$ following \CIGVO.
N-point correlation functions can be easily
generalized.
We insert a state $|\Psi \>$ at $P$ on $\Sigmah$
as shown in figure 2.  Instead of using the projection ${\hat \pi}$ as in
\AGGMV, we {\it choose} a section $\sigma : \Mhg1 \rightarrow \Phg1$.
A volume density $\Os$ on $\Mhg1$ is related to an integration density
$\Ot$ on $\Phg1$ by the pullback, $\Os = \sigma^* \Ot$ or
\eqn\epullb{\Os (V_1,\ldots,V_{3g-2},
                 \Upsilon_1,\ldots,\Upsilon_{3g-2}) =
            \Ot (\vt1,\ldots,{\tilde v}_{3g-2},
                 \nuts1,\ldots,{\tilde \nu}_{3g-2})}
where $\vt{i}=\sigma_*V_i$ and ${\tilde \nu}_i =\sigma_* \Upsilon_i$,
$i=1,\ldots,3g-2$ are linearly independent even and odd vectors respectively
and
$({\vec V},{\vec \Upsilon})$ span the full $T\Mhg1$.
All complex conjugates are suppressed.

Let $\zp$ be the coordinate centered at the chosen point $P$.
We define $\Ot$ on $\Phg1$ by \AGGNSV\dil
\eqn\emeasP{\eqalign{ \Ot &(\vt1,\ldots,{\tilde v}_{3g-2},
                 \nuts1,\ldots,{\tilde \nu}_{3g-2}) \cr
&\equiv \langle \Sigmah,\zp | B[\vt1] \ldots B[{\tilde v}_{3g-2}]
      \delta(B[\nuts1]) \ldots \delta(B[{\tilde \nu}_{3g-2}])
       |\Psi \>^P}}
where $B[\tv=\tv^z \pa_z + {\bar \tv}^{\bar z}\pa_{\bar z}]
= \oint [dzd\theta d\xi]B_z \tv^z
+ \oint [d{\bar z}d{\bar \theta} d{\bar \xi}]{\bar B}_{\bar z} \tv^{\bar z}$,
$B[\nut = \nut^z \pa_z] = \oint [dz d\theta d\xi]B_z \nut^z$ and
similarly its conjugate.
$B_z$ is the antighost superfield in \eghosts.
These definitions can be simplified
by integrating out $\theta$ and $\xi$, but before doing so,
we need to know the following.
When we mod out ${\cal D}_-$, a holomorphic even tensor field \evtz\
becomes ${\tilde v}^z=v_0^z + \theta \omega^\xi$, but given the latter,
we can uniquely lift it back.
Hence, lifting the augmented $\tv
\in T\Phg1$
to
$\tv=(v_0^z +\theta\omega^\xi + \theta\xi \pa_z v_0^z) \pa_z
+({\bar v}_0^{\bar z} +{\bar\theta}{\bar \omega}^{\bar \xi}
+ {\bar\theta}{\bar \xi} \pa_{\bar z} {\bar v}_0^{\bar z}) \pa_{\bar z}$,
we obtain
\eqn\eBve {B[\tv]=- \oint dz (\beta \omega^\xi + b v_0^z)
-\oint d{\bar z}
({\bar \beta} {\bar \omega}^{\bar \xi} + {\bar b} {\bar v}_0^{\bar z}) .}
Since an odd vector $\nut= \theta {\tilde w}^\xi \pa_z \in T\Phg1$ is always
proportional to $\theta$ in the augmented construction (see \epushfo),
we get the very simple form
\eqn\eBvo{ B[\nut] = -\oint dz \beta {\tilde w}^\xi}
and similarly its conjugate.

The $B$ insertions are required also to absorb the fermionic and
bosonic zero modes of $b$'s and $\beta$'s respectively.
The state $\langle \Sigmah, \zp|$ has an anomaly associated with
ghost charges $(U_{b c}, U_{\beta,\gamma}) = (3g-3,3g-3)$, and the
antighosts $b$ and $\beta$ have ghost charges $(-1,0)$ and $(0,-1)$,
the inserted state $\Psi$'s (and in general $N$ inserted states')
ghost charges
must add to yield a net zero ghost number for $\Ot$ in order to have a
nonzero answer.

We can generalize the above prescription slightly \AGGMV.  Suppose we
have a family of surfaces with more than one puncture.  We then can
(and will later) consider families of local coordinates where several
$\sigma_{P_i}$ all depend on the same modulus $m^a$.  Then
$\sigma_*(\pa_{m^a})$ will involve Schiffer variations of
$(\Sigma,\sigma_{P_1},\ldots,\sigma_{P_N})$ at several points, that
is,
\eqn\epushfs{\sigma_*(\pa_{m^a}) = i_{\sigma_{P_1}}(v_a^{(1)}) +
\ldots + i_{\sigma_{P_N}}(v_a^{(N)}) }
where $v_a^{(i)}$ acts at $P_i$.
In this case we replace insertions like \eBve\ and \eBvo\ with the sum
of the corresponding insertions at $P_i$ using $v_a^{(i)}$.

Since we have chosen a section $\sigma : \Mhg1 \rightarrow \Phg1$, there
is no ambiguity in pushing forward the given vectors $({\vec V},{\vec
\Upsilon})$
from $T\Mhg1$ to $(\vt{i},\nuts{i}) \in T\Phg1$ by $\sigma_*$.
However, the correlation function $\int_{\Mhg1} \Os$ now has an apparent
dependence on what section is being chosen.  Consider a nearby
slice $\sigma '$. Then to eliminate this dependence, we impose the
condition $\int_{\Mhg1} \Os - \Omega_{\sigma '} = 0$.
Substituting $\Os =\sigma^* \Ot$, we have
\eqn\eclosed{\eqalign{ 0= \int_{\Mhg1} {(\sigma - \sigma ')}^*\Ot
  =\int_{\pa K} \Ot \cr = \int_K d\Ot ,}}
where $K$ is the enclosed volume between $\sigma$ and $\sigma '$, and
the superspace form of Stokes' theorem \Manin\ is used in the last
step.  This analysis is
correct for the case when $K$ stays away from the boundaries of $\Mhg1$,
otherwise, see below.
Since the space $K$ is arbitrary, we therefore must demand that $\Ot$
to be a closed form $d\Ot =0$.
We have
a closed $\Ot$ only if the state $|\Psi \>$ inserted is
BRST closed, that is $Q_T |\Psi \> =0$ \AGGMV.
Thus for $\int \Os$ to be $\sigma$ independent, we must at least impose the
condition that all inserted states be BRST closed.
Since $d\Ot =0$, it follows that $d\Os = \sigma^* d\Ot =0$.
Therefore, $\Os$ is a closed form on $\Mhg1$.

It seems that the only condition on the states is that they are
BRST closed.  However, the above analysis assumes the existence of a
global slice $\sigma$.  Recall that a point in $\sigma$ is given by
$(\Sigmah ,P,\zp)$.  We will discuss the situation with an ordinary \RS\
$\Sigma$ and then obtain the results on a semirigid surface $\Sigmah$
by the method of augmentation.
Certainly, we can have a coordinate $\zp$ centered at $P$
when the point $P$ is chosen on ${\it any}$ Riemann surface.
However, for $g \not= 1$ Riemann surface, there is a topological
obstruction to have a
smoothly varying {\it family} of local
coordinate systems $\zp$ when $P$
spans the entire surface.
This topological obstruction to having a global coordinate on a compact
$\Sigma$ is given by the Euler number of the manifold.

Let us elaborate on this point.
We cover the ordinary moduli space $\Mg1$ with a family of local
coordinate patches.  On a patch $\alpha$, recall that the
coordinate $z_P^\alpha(Q)$ centered at $P$ is a holomorphic function
of $Q$, but not necessary of $P$.  On an overlap of two patches with
transition given by $z_P^\alpha(Q) = M^{\alpha\beta}(P) z_P^\beta(Q)$,
$M^{\alpha\beta}$ is a $P$-dependent element of the complexified Virasoro
semigroup \Segal.  But the latter is topologically equivalent to
the U(1) group generated by $l_0 -{\bar l_0}$.  So we may deform
these families of coordinates until they agree up to functions with
values in $U(1)$.  An explicit construction of a global family of
coordinates modulo $U(1)$ phases was given by Polchinski \JPFBSA.

A similar analysis for the family of semirigid coordinates can be
carried out.  We begin with a family of ordinary \RS s
with local holomorphic coordinates.  As argued above, we can choose families of
coordinates so that the transition functions are $U(1)$-valued.  We
can then augment the coordinate families as well as families of \RS s
by the procedure given in \eamod.  We must choose however to augment
${\vec m}$ (which include the $3g-3$ moduli and the modulus associated
to $P$) to ${\vec m}+\theta {\vec \zeta}$ but leave ${\vec{\bar m}}$
alone so that we get
$$z_P^\alpha(Q) = exp[2\pi i f_{\alpha \beta}( {\vec m}
+\theta {\vec \zeta}, {\vec{\bar m}})]  z_P^\beta(Q).  $$
We have to leave ${\vec{\bar m}}$ alone because
augmenting them yield ${\bar \theta}={\bar \theta}(Q)$ which is
not a holomorphic function of $Q$.  This transition function
between the two semirigid coordinates is not a pure phase nor is it
$Q$-independent.  However, it differs from a pure phase by a
factor of $[1+\theta(Q)\zeta^a\pa_{m^a} f_{\alpha \beta}( {\vec m},
{\vec{\bar m}})]$.  The one forms  $\{dm^a\pa_{m^a} f_{\alpha \beta}\}$
amount to a 1-cocycle of smooth sections of the holomorphic cotangent
to the moduli space.  Since the associated ${\check {\rm C}}$ech
cohomology vanishes, we may finally modify the coordinates
$z_P^\alpha(Q)$ on each patch to eliminate these factors \PNpc.
\foot{In other words, since a coordinate centered at $P$ is equivalent
to another
if they differ by an even invertible function,
we can redefine the local coordinate at each $P$ to absorb the
unwanted invertible factor.
}
Hence one can choose a family of local semirigid coordinates so that
on each overlap, they differ by a pure phase,
\eqn\emscno{z_P^\alpha (\cdot)=e^{2\pi i f_{\alpha\beta}(P)} z_P^\beta
(\cdot).}
Thus, we have only to ensure that $\Ot$ is insensitive to this phase so that
$\Os=\sigma^* \Ot$ is a well defined measure on $\Mhg1$.

Locally on the overlap of two coordinate patches differing by an infinitesimal
phase, $\sigma'=\sigma+i\epsilon \sigma$ where $\sigma=\zp$ of the
above discussion
(note that it is not $\sigma'=\sigma+i\theta \rho \sigma$),
there is thus a remaining ambiguity in lifting the vectors
$(V_i,\Upsilon_i) \in T\Mhg1$
to $(\sigma_* V_i,\sigma_* \Upsilon)$ or to $({\sigma '}_* V_i,{\sigma
'}_* \Upsilon_i)$.
Imposing in \emeasP\ that $\Ot$ be insensitive to this lifting,
$\delta \Ot (\vt1,\ldots, \tv_{3g-2},
                 \nuts1,\ldots,\nut_{3g-2})=0$,
we obtain the condition ${(b_0 -{\bar b}_0)}|\Psi\> =0$ \dil.
Moreover, there are coordinate dependences on the state via \estatem.
Imposing the condition that this dependence also drop out and by using
\ealgebra, it yields $(L_0 - {\bar L}_0)|\Psi \>=0$.
Stronger conditions on the state
that save algebra without sacrificing
any interesting observables are
\eqn\eequiv{L_0 \sPsi = b_0 \sPsi = Q_T \sPsi = 0 \quad
{\rm and~their~conjugates.}}
These are known as the equivariance or weak physical state conditions
(WPSC) on $\sPsi$ \TCCT.  Note that there is no such condition imposed by
$G$ or $\beta$ on $\sPsi$.

We have ignored the boundaries of moduli space in the above discussion.
However, they are important because they contribute to the correlation
function as contact terms.
If there are boundaries in the moduli space, then
\eclosed\ is not valid,
but if we specify boundary conditions on $\sigma$, then we may recover
a well defined integral $\int \Os$ over the moduli space.
To get such boundary conditions, note that $\Mhg{N}$ is non-compact, but
near a boundary, there exists a notion of good coordinates, those
which are compatible with
the stable-curve compactification of moduli space \semirig.
Hence, we can specify what the asymptotic  slices are at the relevant
boundaries
and thus get a well defined measure over the entire $\Mhg{N}$.
They are given by \egsc\ near a boundary of $\Mhg{N}$ when two punctures
approach each other.
Moreover, we will see in section 5 that the choices
$(a_i,E)$ and $({\tilde a}_i, {\tilde E})$ of the asymptotic slices \egsc\
drop out at the end of the calculation.

We will now give the relation between equivariant BRST cohomology and
deRham cohomology.  Here, we are not interested in the
moduli space's boundary contributions,
hence we will assume that
$\Mhg{N}$ is a compact manifold without boundaries.  First, consider a
closed n-form $\Omega$ defined
on a compact \RS\ $M$.
If $\Omega = d\mu$, where $\mu$ is an $({\rm n}-1)$-form,
then we call $\Omega$ exact.
The deRahm cohomology is defined as the vector space $H^n(M)$,
where $H^n(M) \equiv {\rm closed~n~forms / exact~n~forms}$.
That is, if $H^n(M) \not= 0$, then a closed n-form need not be exact.
In fact, $\int_M \Omega$ depends only on the cohomology class of
$\Omega$ in $H^n(M)$.
For a BRST-exact state, $\sPsi = (Q_T + {\bar Q}_T)|\lambda \>$, one can show
that the closed form $\Omega_\Psi = d\mu_\lambda$ is d-exact \AGGMV\LaNelson;
then $\int_M d\mu_\lambda =0$ because $M$ is compact.  In other words,
BRST-exact states decouple.  This discussion can be generalized to the
top density $\Os$ on $\Mhg1$ we obtained from \epullb.  Then, integrating
over the fiber of $\Pi : \Mhg1 \rightarrow \Mg1$, $\int_{\Pi^{-1}} \Os$
yields the closed
form $\Omega$ of the above discussion
where the manifold $M$ becomes $\Mg1$.

In the above argument, we need to keep in mind the equivariance
condition or WPSC
\eequiv.
In fact in topological gravity all BRST-closed states are also
BRST-exact \VVTG,
except for the states created by the puncture operator and the vacuum state.
The other non-trivial
observables in topological gravity can therefore all be
written as \VVTG \TCCT
\eqn\eQmu{\sPsi = (Q_T+ {\bar Q}_T) |\lambda\>, \quad {\rm where}~|\lambda\>
{}~{\rm fails~the~WPSC}.}
It is convenient to break the Hilbert space of BRST-closed states
into two disjoint sectors,
${\cal H}^{\rm BRST-closed} = {\cal H}^{\rm WPS} \oplus
{\cal H}^{\rm rest}$, so that $\sPsi \in {\cal H}^{\rm WPS}$
and $|\lambda \> \in {\cal H}^{\rm rest}$.
However, this does not mean that the theory is empty.
A state like \eQmu\ need not decouple because it is not really a
BRST-exact state in the equivariant sense, so that $\mu_\lambda$ in
the previous paragraph is not globally defined on $\Mg1$.

For an n-form $\Omega$ on a compact manifold $M$ where the ${\rm
n^{th}}$-cohomology
is non-trivial $H^n(M) \not= 0$,
by the Poincar\'e lemma we can always write $\Omega = d\mu_i$ on a
local patch $U_i$, $\cup U_i = M$. Then
\eqn\estokes{\int_M \Omega = \sum_i \int_{U_i} d\mu_i =
           \sum_i \int_{\pa U_i} \mu_i ~.}
If the $\mu_i$ agree along the boundaries $\pa U_i$, then $\Omega$
is an exact form globally on $M$ and $\int_M \Omega = 0$.
Generalizing to our semirigid situation, we insert the state \eQmu\
into $\Ot$ of
\emeasP and obtain
$\Ot_\Psi = d \mut_\lambda$, where d is the ordinary exterior
derivative and $\mut_\lambda$ is an integration density of degree
$(3g-3,3g-2)$.
Total derivatives in the Grassmann variables are ignored because they
vanish when integrated.
Observe that $\mut_\lambda$ will not
be invariant
under the change in phase in the coordinates across patches
because the state $|\lambda\>$ does not satisfy the WPSC.
What this means is that on a local patch on the moduli space, we have
by \epullb\ $\Omega_\sigma |_\alpha = d (\sigma^*|_\alpha \mut_\lambda)$,
and on an
overlapping patch,  $\Omega_\sigma |_\beta = d (\sigma^*|_\beta \mut_\lambda)$,
but $\sigma^*|_\alpha \mut_\lambda$ and $\sigma^*|_\beta \mut_\lambda$
do not agree on the overlap.
Just like the discussion above, since $\mu_i= \int_{\Pi^{-1}} \sigma^*|_i
\mut_\lambda$
(the integral over the fiber of $\Pi : \Mhg1 \rightarrow \Mg1$)
do not agree on the
overlap, the top form $\int_{\Pi^{-1}} \Os$ can give a nontrivial element in
$H^{6g-4}(\Mg1)$. Therefore the correlation function
$\int \Os = \int \sigma^* \Ot_\Psi$ probes non-trivial
topology of the moduli space $\Mg1$ of the family of \RS s.

The non-trivial observables were given by E. and H. Verlinde \VVTG.
We will in this paper use the modified version of the non-trivial observables
given by Distler and Nelson in \dil, and further, we normalize them by the
factor $1/2\pi i$.  They are for $n \ge 0$,
\eqn\eobserv{|\cO{n} \> = {1 \over 2\pi i} {\gamma_0}^n c_1 {\bar c_1}
|-1\> , \quad |-1\> \equiv \delta (\gamma_1) \delta ({\bar \gamma_1}) |0\>,
}
where $|-1\>$ denotes the vacuum state at Bose sea level $-1$.  These
observables $|\cO{n} \>$ satisfy the WPSC \eequiv.  However, these states are
non-trivial because when written as BRST-exact states using
\ealgebra, for $n \ge 1$,
\eqn\eOQmu{ | \cO{n} \> = (Q_T+{\bar Q}_T) | \lambda_n\> , \quad |\lambda_n \>
= c_0
| {\cal O}_{n-1} \> ,}
$|\lambda_n\>$ fails the WPSC since $b_0|\lambda_n\> \not= 0$ although it
satisfies the rest of the WPSC.  This fits the category of states
\eQmu.  That is, these states
are BRST-closed but nonetheless not BRST-exact in the equivariant sense.
In particular, $\cO0$ is called the puncture operator and $\cO1$ the dilaton.
The puncture operator is the unique operator that satisfies stronger
conditions than WPSC, namely the strong physical state condition
\dil\TCCT : $Q_T |\cO0\> = L_n|\cO0\> = G_n |\cO0\> = b_n |\cO0\> =
\beta_n |\cO0\> =0$ for $n \ge 0$ and their complex conjugates.
Note that $|{\cal O}_{n \ge 1} \>$
would have been a strong physical state if
it were not for $\beta_0 |{\cal O}_n \> =0$.
Although $\cO0$
is a BRST-closed state it cannot
be written as BRST-exact state, hence it need not decouple.

By conservation of ghost charges $(U_{bc}, U_{\beta \gamma})$, one can
show that for a correlation of $N$ observables $\cO{n_i}$ to be
nonzero, the number $N$ and type $n_i$ of
observables \eobserv\ are constrained by $\sum_{i=1}^N n_i = 3g-3 + N$ \dil.
In the remaining sections we will assume that this constraint is
satisfied
by having the correct amount and type of observables.  (Sometimes we
will not show all of them explicitly.)

\newsec{The dilaton equation}

\def\zp{z_P{(\cdot)}}
\def\zr{z_r{(\cdot)}}
\def\zpbar{{\bar z}_p{(\cdot)}}
\def\isig{ i_\sigma}

In this section, we will integrate over the location of the dilaton
first and stay away from the boundaries of the moduli space.
The boundary contributions giving contact terms are considered in \dil.
We will suppress all 
moduli from the formulas except the $1|1$ moduli
associated to the chosen puncture $P$.
We will also suppress other inserted states needed for ghost
charge conservation and display only the dilaton.
In particular, the zero mode insertions for the $3g-3|3g-3$ moduli
associated to the unpunctured surface are inserted at punctures
other than $P$.  This is allowed by the operator formalism \AGGMV.

To begin, we take an ordinary \RS\ and choose a covering by coordinate
patches and
a corresponding tiling by polygons with edges along two-fold patch
overlaps and vertices at three-fold overlaps.  For example, we can
choose in such a way that each polygon is a triangle.  We then augment
this \RS\ to a semirigid surface with semirigid coordinate patches.
On the overlap of two patches,  recall that their respective
coordinates can be chosen to differ by a $U(1)$ phase as in \emscno.
This prompts us to define a semirigid coordinate family to include a possible
$U(1)$ phase $M(P)$, $\zp \equiv M(P) \sigma_P(\cdot)$ on a local
patch where $\sigma_P(\cdot)$ is a general coordinate.
When $P$ is on the overlap with another patch, $\sigma_P(\cdot)$ is
chosen to be common to the two
overlapping patches whereas the phase $M(P)$
jumps across patches.
To keep our calculations simple, instead of the general coordinate
$\sigma_P(\cdot)$, we will illustrate with the ``conformal normal
ordered'' \JPvert\LaNelson\
coordinate $z-(r+\theta\rho)$ where $(r,\rho)$ are the moduli
associated to $P$. As we will see, it is only the phase difference
across patches that
matters and hence we will obtain the same answer if we begin with a general
coordinate $\sigma_P(\cdot)$.
Thus, on a local patch $U_\alpha$ 
the modified conformal normal ordered coordinate
\eqn\emcno{\zp = M(P)(z-r-\theta\rho)}
will be used.
Recall that the phase $M(P)$ by construction depends on the even
moduli $r$ and ${\bar r}$ but not their odd partners associated to the
location of $P$
and jumps across coordinate patches.

We will first find the vectors
$({\tilde v}_i = \sigma_* V_i, \nuts{i} = \sigma_* \Upsilon_i)
\in T{\cal P}$ that deform the moduli $(r,\rho)$ associated to $P$
on the coordinate patch $U_\alpha$ and the corresponding ghost
insertions for the measure.  Here, we have
$(V_{3g-2},\Upsilon_{3g-2}) =(\pa_r, \pa_\rho)$ and their complex
conjugates $(\pa_{\bar r}, \pa_{\bar \rho})$.
Using \emcno, we get
\eqn\epfr{\eqalign { \sigma_* ( { \pa \over \pa r} ) &=
{ { \pa \zp } \over { \pa r}} { \pa~ \over { \pa \zp}}
+ { { \pa \zpbar} \over {\pa r} } { \pa~ \over { \pa \zpbar} } \cr
 &= - M { \pa~ \over { \pa \zp} } + A \zp { \pa~ \over { \pa \zp} }
+ \bar B \zpbar { \pa~ \over { \pa \zpbar} } \cr
 &= M \isig (l_{- 1}) - A \isig (l_0)
- \bar B \isig ( \bar l_0) } }
where
${ A \equiv M^{-1} \pa_r M ~,~
\bar B \equiv {\bar M}^{-1} \pa_r \bar M}$ and $\sigma=\zp$.
Similarly,
\eqn\epfs{\eqalign{ \sigma_* ({ \pa \over {\pa \bar r}} ) =
\bar M \isig ({\bar l}_{-1}) - \bar A \isig ({\bar l}_0)
- B \isig (l_0) , \cr
\sigma_* ({ \pa \over {\pa \rho}}) = - M \isig (g_{-1}),
 \quad {\rm and} \quad
\sigma_* ({ \pa \over {\pa \bar \rho}}) =
- \bar M \isig ({\bar g}_{-1}).}}
According to \eBve\ and \eBvo, therefore
\eqn\eBh{\eqalign { B  \lbrack{\pa \over {\pa r}}\rbrack   =
M b_{-1} - A b_0
- \bar B {\bar b}_0 \equiv {\hat b}_{-1} , \quad
 B  \lbrack{\pa \over {\pa \bar r}}\rbrack   = \bar M b_{-1}
 - \bar A {\bar b}_0
- B b_0 \equiv {\hat {\bar b}}_{-1} ,\cr
 B  \lbrack{\pa \over {\pa \rho}} \rbrack   = -M \beta_{-1} \equiv
-{\hat \beta}_{-1} , \quad {\rm and} \quad
  B  \lbrack{\pa \over {\pa \bar \rho}}\rbrack  =
 - \bar M {\bar \beta}_{-1}
\equiv -{\hat {\bar \beta}}_{-1} .   } }
On another overlapping patch $U_\beta$ in which a coordinate
with the phase $\tilde M$ is used, we let
\eqn\eMt{ M = e^{2 \pi i f} {\tilde M} }
to be the transition map, where $f$ is real and depends
only on $r$ and ${\bar r}$.  Thus on the overlap between these two patches,
we have
\eqn\eABt{ {\tilde A} - A = - 2 \pi i \pa_r f  \quad {\rm and} \quad
{\tilde B} - B = - 2 \pi i \pa_{\bar r} f. }

{}From the discussion in section 4, if we have a measure
$\Ot_{\Psi=Q\lambda}$, then
\eqn\estokes{ \int \Ot_{Q\lambda} = \sum_i \int_{U_i} d \mut_\lambda |_i
= \sum_i \int_{\pa U_i} \mut_\lambda |_i, }
where $\mut_\lambda |_i$ means $\mut_\lambda$ evaluated in the coordinate
patch $U_i$.
On the patch $U_\alpha$, using \eBh, the boundary contributions
give
\eqn\edila{\eqalign { \int_{\pa U_\alpha} \mut_\lambda |_\alpha
= \int_{\pa U_\alpha} dr d^2 \rho \< \Sigma , \zp | {\hat b}_{-1}
\delta ( {\hat \beta}_{-1}) \delta ({\hat {\bar \beta}}_{-1})
| \lambda \>  \cr  +
\int_{\pa U_\alpha} d{\bar r} d^2 \rho \< \Sigma , \zp | {\hat {\bar b}}_{-1}
\delta ( {\hat \beta}_{-1}) \delta ({\hat {\bar \beta}}_{-1})
| \lambda \> }}
and on the patch $U_\beta$, we let $M$ in \edila\ go to
$\tilde M$.
The sum in \estokes\ can be rearranged, replacing integrals along $\pa
U_i$ by integrals along the common edges $W_{\alpha \beta}$ on the overlaps
$U_\alpha \cap U_\beta$ as follows.
\eqn\edilr{\eqalign{ \int \Ot_{Q \lambda} &= \sum_{(\alpha,\beta)}
\int_{\pa U_\alpha \cap W_{\alpha \beta}} \mut_\lambda |_\alpha
+ \int_{\pa U_\beta \cap W_{\alpha \beta}} \mut_\lambda |_\beta \cr
&= \sum_{(\alpha,\beta)} \int_{W_{\alpha \beta}} \mut_\lambda |_\alpha -
\mut_\lambda |_\beta}}
where $(\alpha,\beta)$ denote that the sum is over each overlap
$W_{\alpha \beta}$ once.  The sign change in the last step is due to
reversing the orientation of the line integral.

To evaluate \edila, we first integrate over the odd moduli
$\rho$, so we need to extract the $\rho$ dependence
in $\<\Sigma, \zp | $. That is from \estatem,
\eqn\estateG{\eqalign{ &\< \Sigma , \zp = M(z - r - \theta \rho) |\cr
&= \< \Sigma , \zr = M (z - r) | \quad ( 1 - \rho M G_{-1} )
( 1 - \bar \rho \bar M {\bar G}_{-1}).}}
Since there are no other $\rho$ dependencies in the rest of
the measure, integrating over $\rho$ give us the picture changing
operators, and we are left with
\eqn\edilab{\eqalign{ \int_{\pa U_\alpha} \mut_\lambda |_\alpha =
\int_{\pa U_\alpha} dr
\< \Sigma , \zr | {\hat b}_{-1} G_{-1} \delta (\beta_{-1})
{\bar G}_{-1} \delta ({\bar \beta}_{-1}) | \lambda \> \cr +
\int_{\pa U_\alpha} d \bar r
\< \Sigma , \zr | {\hat {\bar b}}_{-1} G_{-1} \delta (\beta_{-1})
{\bar G}_{-1} \delta ({\bar \beta}_{-1}) | \lambda \> }}
where
$ \delta({\hat \beta}_{-1}) ={ |M|^{-1} \delta( \beta_{-1} )}$
and its complex conjugate are used.\break\hfil
Since
$| {\cal O}_1\>  = (Q_T + {\bar Q}_T)|\lambda\>$ where
$|\lambda\> = (2\pi i)^{-1} c_0 c_1 {\bar c}_1 | -1 \>$
and \break\hfil
$G_{-1} \delta (\beta_{-1}) {\bar G}_{-1}
\delta ({\bar \beta}_{-1}) c_1 {\bar c}_1 |-1\> = -|0\>$,
we obtain
\eqn\edilac{
(2\pi i)~\int_{\pa U_\alpha} \mut_\lambda |_\alpha
=  - \int_{\pa U_\alpha} dr \< \Sigma , \zr | {\hat b}_{-1}
c_0 | 0 \> - \int_{\pa U_\alpha} d \bar r \< \Sigma , \zr | {\hat
{\bar b}}_{-1}
c_0 | 0 \>. }
Using \eBh, the only non-vanishing term is
\eqn\edilad{
(2\pi i ) \int_{\pa U_\alpha} \mut_\lambda |_\alpha
= ( \int_{\pa U_\alpha} d r A
+  \int_{\pa U_\alpha} d \bar r B \quad) Z }
where
$ Z = \<\Sigma, \zp | 0\> $
along with the suppressed
$(3g-3, 3g-3)$ ghost insertions.
Thus by \edilr,
\eqn\edilfr{\eqalign  { \<{\cal O}_1 \ldots\> &= \sum_{(\alpha , \beta)}
(2\pi i )^{-1} ( \int_{W_{\alpha\beta}} \lbrack  d r (A - \tilde A)
+ d \bar r (B - \tilde B)  \rbrack ) Z   \cr
&= \sum_{(\alpha , \beta)}
( \int_{W_{\alpha\beta}} \lbrack  d r \pa_r f
+ d \bar r \pa_{\bar r} f \rbrack  ) ~ Z  \cr
&=  \chi Z , \quad {\rm where} \quad \chi = 2g-2. }  }
The last step is given by the standard definition of the Euler number.
As mentioned we can tie on the calculation of \dil\ to get the full
dilaton equation \eDE.

\newsec{The puncture equation}
The expression \ePE\ can be derived by integrating out the puncture
operator $\cO0$ from the lhs of \ePE.
We will first show that the puncture
operator, unlike the dilaton, gives no contribution when integrated
over the bulk of the moduli space.  The contributions come only near
the boundaries, when the puncture runs into other inserted states.
We will then analyze a contact term and show that
$\< \cO0 \cO{n}\ldots \> = n \< \cO{n-1}\ldots \>$
which can be generalized
to the case when there are $N$ contact terms,
$\< \cO0 \cO{{n_1}} \ldots \cO{{n_N}} \>$.

To begin, we ignore the boundaries of the moduli space and use the family
of modified conformal normal ordered coordinates \emscno\ on the bulk
of moduli space
$\Mhg1$. \foot{Since the puncture operator satisfies the strong physical state
conditions, it does not matter what local coordinate slice we choose to
evaluate its contribution in the bulk of moduli space.
We may as well use the family of conformal normal ordered
coordinates.  However, we will use the family of modified conformal
normal ordered coordinates and see explicitly that the phase $M$ is
irrelevant.}
The correlation function is given by
\eqn\ePOn{\eqalign{ &\< \cO0 \prod_{i=1}^N \cO{{n_i}} \> =
\int d^2r d^2\rho \< \Sigmah, \sp, \ldots| \cr
&\times B[\sp_*(\pa_r)] B[\sp_*(\pa_{\bar r})]
\delta ( B[\sp_*(\pa_\rho)]) \delta ( B[\sp_*(\pa_{\bar \rho})])
|\cO0 \>^P \otimes_{i=1}^N |\cO{{n_i}} \>,}}
where $(r,\rho)$ are the parameters (moduli) describing the position $P$
of the puncture operator.  $|B[\sp_*(\pa_r)] \delta ( B[\sp_*(\pa_\rho)])
|^2$ is the zero mode insertion associated to the point $P$.
The $(3g-3+N,3g-3+N)$ other zero mode insertions are suppressed from notation
and inserted
elsewhere away from $P$. 
Thus, we only display terms that depend on $(r,\rho)$ which are
going to be integrated out.

On a coordinate patch $U_\alpha$,
we use \emcno\ for $\sp$ and by substituting its push forward \eBh\ into
\ePOn, we obtain
\eqn\ePOna{\eqalign{ &\< \cO0 \prod_{i=1}^N \cO{{n_i}} \> = \cr
&\int d^2r d^2\rho  \< \Sigmah, M(z-r-\theta\rho), \ldots|
{\hat b_{-1}^P} {\hat {\bar b}}_{-1}^P
\delta({\hat \beta_{-1}^P}) \delta({\hat {\bar \beta}}_{-1}^P)
|\cO0 \>^P \otimes_{i=1}^N |\cO{n_i} \>,}}
where the integral over other moduli are again implicit.
Substituting \eBh\ and \eobserv, we have
${\hat b_{-1}^P} {\hat {\bar b}}_{-1}^P \delta({\hat \beta_{-1}^P})
\delta({\hat {\bar \beta}}_{-1}^P) |\cO0 \>^P = -|0 \>^P$.
Next, integrating over the fibers $\Pi : \Mhg1 \rightarrow \Mg1$ implies
integrating over $\rho$.  Since there are no dependences on $\rho$
other than the state $\< \Sigmah, M(z-r-\theta\rho), \ldots| $,
this integration brings out the factor $|M|^2 {\bar G_{-1}^P} G_{-1}^P$
using \estateG.   Since $|0\>$ is an augmented-$SL(2,{\spec C})$
invariant vacuum,
$G_{-1}^P |0\>^P=0$. Thus on each coordinate patch
$\< \cO0 \prod_{i=1}^N \cO{{n_i}} \>$ vanishes; this correlation
function does not get contributions from the bulk of moduli space.

At a boundary of the moduli space when the puncture runs into an
inserted state, the modified conformal normal ordered coordinate near
$P$ is no longer allowed. We have to use the coordinate slice \egsc\
obtained from the sewing prescription.
Suppressing other operators, a contact term between the puncture and
$\cO{n}$ is given by
\eqn\econtact{\eqalign{\< \cO0  \cO{n} \> =
\int d^2q d^2\delta \< \Sigmah, \sp , \sigma_Q|
&B[\sigma^{PQ}_* (\pa_q)] B[\sigma^{PQ}_* (\pa_{\bar q})] \cr \times
&\delta ( B[\sigma^{PQ}_* (\pa_\delta)])
\delta ( B[\sigma^{PQ}_*(\pa_{\bar \delta})])
|\cO0 \>^P \otimes |\cO{n} \>^Q.}}
$(q,\delta)$ is the position of the point $P$ {\it relative} to $Q$,
and
$\sigma^{PQ}_*= \sigma_{P~*} \oplus \sigma_{Q~*}$. The push forward of a
vector by
$\sigma^{PQ}_*$ lives in the vector space $T\Phg2$ \dil\ which
is the generalization of $T\Phg1$.
$\<\Sigmah, \sp,\sigma_Q|$ is a state in the dual of the tensor
product of two copies of Hilbert space.
The two copies of Hilbert space
consist of states labelled by $P$ and $Q$ respectively.

When $q \approx \epsilon$ is small and given that $\sigma$
is the coordinate on $\Sigmah$ centered at
the attachment point to the standard three punctured sphere, the
sewing prescription yields the general coordinate \egsc\ centered at $P$
$$ {\tilde \sigma}_P = \Sp + a_1 \Sp^2 + a_2 \Sp^3 + \ldots
\quad {\rm where} \quad
\Sp = {\sigma \over q} - E - {\theta \delta \over q^2}\sigma ,$$
and similarly for the coordinate centered at $Q$
with $E \rightarrow {\tilde E}$ and $a_i \rightarrow {\tilde a}_i$.
When $P$ and $Q$ are far apart, $q \gg \epsilon$, we should interpolate
to the conformal
normal ordered slice $\sp^{c.n.o.} = \sigma - qE - \theta\delta E$
up to a phase.  However, we have imposed a stronger condition WPSC
\eequiv\ than the necessary phase independent condition; the density
$\Ot$ is insensitive to changes in the section $\sigma$ by complex
multiplicative factors.  Hence, the section is globally defined modulo
a complex multiplicative factor.  This saves algebra because we can
smoothly interpolate between ${\tilde \sigma}$ for $|q| < \epsilon$
and $q^{-1}\sp^{c.n.o.}$ for $|q| > 2\epsilon$ by
an interpolating function $f(|q|)$ as in \TCCT, where
$f=0$ for $|q| < \epsilon$ and
$f \rightarrow 1$ as $|q| \rightarrow \infty$.
Hence the interpolated slice is given by
\eqn\eints{{\check \sigma}_P = \Sp + A_1 \Sp^2 + {f \theta \delta \over q}\Sp
+\ldots}
where the ellipsis are terms involving $a_{n \ge 2}$ and
$A_1 = a_1 (1-f)$.  We have a similar expression for the interpolated
slice at $Q$ by letting the parameters become their tildes.
The integrand has two parts, the state
$\< \Sigmah, {\check \sigma}_P, {\check \sigma_Q}|$ and the zero mode
insertions $|B[(\pa_q)] \delta ( B[(\pa_\delta)])|^2$.
Their $(q,\delta)$ dependences will be extracted independently,
combined and integrated over, $\delta$ first and then $q$.

By using \estatem, we extract those possibly contributing $(q,\delta)$
moduli dependences from
$\< \Sigmah, {\check \sigma}_P, {\check \sigma_Q}|$.
To do this, we turn
${\check \sigma}_P$ into composition of maps,
$$ {\check \sigma}_P = (s+ A_1 s^2) + {2A_1 f \delta \theta \over q}
(s + A_1 s^2)^2 + \ldots \quad {\rm where} \quad
s=(1+ q^{-1} f\theta\delta)\Sp.$$
Hence we have
\eqn\ecs{\< {\check \sigma}_P| = \< s|
(1 - A_1  L_1) ( 1-2A_1 f q^{-1} \delta G_1)}
where we have left out terms involving modes $L_{n \ge 2}$ and $G_{n
\ge 2}$ and along with them are $a_{n \ge 2}$.
If the $a_1$ terms do not contribute then neither will the
higher modes as we will see.
Complex conjugates are again suppressed until needed.
Furthermore, we have
\eqn\ess{\eqalign{ \< s| &= \< \Sp|(1+ \delta f q^{-1}G_0) \cr
  &= \< z_E |
{}~(1-\delta E q^{-1} G_{-1} )( 1+ \delta f q^{-1} G_0)  \cr
  &= \< \sigma |~ q^{L_0} e^{EL_{-1}}(1-\delta q^{-1} G_0)
(1-\delta E q^{-1} G_{-1} )( 1+ \delta f q^{-1} G_0) }}
where $z_E = (q^{-1}\sigma -E) - \theta \delta q^{-1}(q^{-1}\sigma -E)$.
$\< {\check \sigma_Q}|$ has the same expansion as $\< {\check
\sigma_P}|$ with $A_1 \rightarrow {\tilde A}_1$ and $E \rightarrow
{\tilde E}$.  We will however set ${\tilde E}=0$ because at the end of
the calculation, we wish that whatever is inserted at $Q$ will be
inserted with the coordinate $\sigma$.
Putting together the expansion in ${\check \sigma}_P$ and ${\check \sigma_Q}$,
we end up with
\eqn\ecczes{\eqalign{&\< \Sigmah, {\check \sigma}_P, {\check \sigma_Q}| \cr
&=\< \Sigmah , z_E, \sigma|\{ (1- \delta E q^{-1} G_{-1}^P)
(1+\delta f q^{-1} G_0^P)
(1 - A_1 L_1^P) ( 1-2A_1 f q^{-1} \delta G_1^P) \} \cr
&\times \{  q^{L_0^Q}(1-\delta q^{-1} G_0^Q) ( 1+ \delta f q^{-1} G_0^Q)
(1 - {\tilde A_1} L_1^Q)
( 1-2{\tilde A_1} f q^{-1} \delta G_1^Q) \}.}}

We are left with giving the zero mode insertions.
They are gotten from the push forwards of $\pa_q$ and $\pa_\delta$ and
their complex conjugates by the
interpolated slices ${\check \sigma_P}$ and ${\check \sigma_Q}$ in \eints.
Setting ${\tilde E}=0$ at $Q$ and keeping only to $a_1$ terms, we obtain
\eqn\ebq{qB[\pa_q]=Eb_{-1}^P + {\bar \delta}(2{\bar q})^{-1} f' |q|
({\bar \beta_0^P}
+ {\bar \beta_0^Q}) + \delta q^{-1} F^{PQ}(E, |q|, a_1),}
where $F^{PQ}(E,|q|,a_1)$ is linear in the operators $\beta^{P,Q}$
and a function of $E$, $|q|$ and $a_1$.
We will later see that this term does
not contribute because it has a $\delta$ coefficient.
Similarly, we have the
conjugation of \ebq\ giving us the push forward of $\pa_{\bar q}$.
We also have
$$ qB[\pa_\delta]= -E\beta_{-1}^P
   -(1-f)(\beta_0^P + \beta_0^Q) +2a_1 E \beta_0^P,$$
thus giving
\eqn\ebd{ \delta( qB[\pa_\delta]) = |qE|^{-1}\delta(\beta_{-1}^P)
+\delta '(E\beta_{-1}^P)\{
(1-f)(\beta_0^P + \beta_0^Q) - 2a_1E\beta_0^P \} }
and similarly its conjugate.
We will first evaluate \econtact\ with only the first term of \ebd\
and with $a_1={\tilde a_1}=0$ which imply $A_1={\tilde A}_1=0$.
It will be shown later that the terms ignored now do not contribute
when turned on.

Substitute \ebq, the first term of \ebd\ and their conjugates into \break\hfil
$B[\pa_q]B[\pa_{\bar q}] \delta(B[\pa_\delta])\delta(B[\pa_{\bar
\delta}]) |\cO0 \>^P \otimes |\cO{n} \>^Q$ of \econtact, we get
\eqn\ebbbb{\eqalign{ &\{Eb_{-1}^P + {\bar \delta}(2{\bar q})^{-1} f' |q|
({\bar \beta_0^P}
+ {\bar \beta_0^Q}) + \delta q^{-1} F^{PQ}             \} \cr
 &\times     \{  E{\bar b}_{-1}^P + {\delta}(2 q)^{-1} f' |q|
({\beta_0^P}
+ {\beta_0^Q}) + {\bar \delta}{\bar q}^{-1}  {\bar F}^{PQ}\}
(2\pi i)^{-1} c_1^P {\bar c}_1^P E^{-2} |0 \>^P \otimes |\cO{n} \>^Q, }}
where $[\delta (\gamma_n),\delta (\beta_m) ] = \delta_{n+m,0}$ is used.
What we want is to integrate out the puncture operator completely
and in \ebbbb\ we have to remove its remnant $c_1^P {\bar c}_1^P$.
First we observe from \ealgebra\ that only the operators $b_{-1}$ and
$G_{-1}$ are conjugate to $c_{1}$.
Without sandwiching with the state
$\<\Sigmah, {\check \sigma_P}, {\check \sigma_Q}|$, we have to take
$E^2 b_{-1}^P {\bar b}_{-1}^P$ to wipe out $c_1^P {\bar c}_1^P$.
However, integrating over $d^2 \delta$ kills it.  Hence we need at
least one $G_{-1}^P$ and/or ${\bar G}_{-1}^P$ from the state since every
$G~({\bar G})$ comes with a $\delta~({\bar \delta})$.
On the other hand if we choose both
$\delta G_{-1}^P~{\bar \delta} {\bar G}_{-1}^P$ from the state, then
picking any terms in $\{~\}$ of \ebbbb\ will give zero because
$\delta^2 = {\bar \delta^2} = 0$ and $G_{-1}|0\>= {\bar G}_{-1}|0\>=0$.
Hence the only possible non-vanishing term has either
$\delta G_{-1}$ or ${\bar \delta}{\bar G}_{-1}$ from the state.
Since we have $\gamma_0^n$ but not its complex conjugate in
$|\cO{n}\>$ in \eobserv, the only contributing combination is
$-{\bar \delta}{\bar q}^{-1} E {\bar G}_{-1}^P$ from the state along
with $Eb_{-1}^P$ and $(2q)^{-1}\delta f' |q| (\beta_0^P + \beta_0^Q)$
from the zero mode insertions in \ebbbb.
Substituting \ecczes\ with $a_1={\tilde a_1}=0$, we finally arrive at
\eqn\ezson{\eqalign{&\< {\cal O}_0 {\cal O}_n \> =\cr
&(2\pi i)^{-1} \int d^2q d^2\delta
\< \Sigmah, z_E, \sigma| q^{L_0^Q} {\bar q}^{{\bar L}_0^Q}
{\bar \delta} \delta  (2|q|)^{-1} f' \beta_0^Q
b_{-1}^P {\bar G}_{-1}^P c_1^P {\bar c}_1^P |0\>^P \otimes |\cO{n}\>^Q}}
since $\beta_0^P|0\>^P =0$.
Using $[L_0, \beta_0]=0, L_0^Q|\cO{n}\>^Q = {\bar L}_0^Q|\cO{n}\>^Q =0$
(WPSC), and $\beta_0^Q |\cO{n} \>= -n|\cO{{n-1}}\>$ for $n \ge 0$, we have
\eqn\ePOnf{\eqalign{\< \cO0 \cO{n} \ldots \>
&=  (2\pi i)^{-1}(-n) \int d^2q (2|q|)^{-1} f'
\<\Sigmah, z_E, \sigma|~|0\>^P \otimes |\cO{{n-1}} \>^Q  \cr
&= n \<\Sigmah, \sigma|\cO{{n-1}}\>^Q \int d|q| f'(|q|) \cr
&= n \<\cO{{n-1}} \ldots \>.}}

Recall that we left out the term with $\delta '(E\beta_{-1}^P)$ in
\ebd.  It requires a $E\beta_{-1}^P$ term so that using
$x\delta '(x)=-\delta (x)$ we get $\delta (E\beta_{-1}^P)$ to raise
the Bose see level back to 0 in $|\cO0 \>^P $.
However, we did not pick up
any $\beta_{-1}^P $ term in integrating out $ d^2 \delta $.
Thus including
$\delta '(E\beta_{-1}^P)$ will not yield new contribution to integrating
out the puncture operator.

Finally, we put back the $a_1$ dependences.
First note that all $a_1$ dependences in the state
$\< \Sigmah, {\check \sigma_P}, {\check \sigma_Q}|$ come with $L_{1}^P$
or $G_{1}^P$ \ecczes.  Since all the operators in
$|B[\pa_q]\delta (B[\pa_\delta ])|^2$
have mode expansion greater than $-1$
and the puncture operator is inserted at $P$, hence by \ealgebra,
$L_{1}^P$ and $G_{1}^P$ will annihilate the state at $P$.
The other $a_1$ dependences come from the zero mode insertions.
$a_1$ comes in via $F^{PQ}$ in $B[\pa_q]$ \ebq\
which did not contribute.
It also comes into $\delta(B[\pa_\delta ])$, appearing with the
the term $\delta '(E\beta_{-1}^P)$ in \ebd\ which we argued will not
contribute.  Thus, turning on $a_1$ does not affect the result
we obtained.
Similarly, $a_{i \ge 2}$ will drop out since they are associated with
higher modes $L_{n \ge 2}$, $G_{n \ge 2}$ and $\beta_{n \ge 1}$.
As for ${\tilde a}_i$, the
dependences come only from the state since we have set ${\tilde E} =0$.
Just like before, they come with $L^Q_{n \ge 1}$ and
$G^Q_{n \ge 1}$. Although these operators do not appear in the
WPSC, they vanish when applied to the state $|\cO{n} \>^Q$ in \ebbbb.
Hence, no ${\tilde a}_i$ terms contribute either.
Thus, the answer is completely independent of the choice of coordinates
around $P$ and $Q$.  More generally, we get a contribution
similar to \ePOnf\ in integrating out the puncture operator
each time it comes in contact with an operator in
$\< \cO0 \cO{n_1} \ldots \cO{n_N}\>$. Hence we finally have the
desired recursion relation \ePE.

\newsec{Conclusions}
In two dimensions, field theory, quantum mechanics and gravity are
compatible.  By imposing the semirigid symmetry, we
simplify the situation enough to see what a theory of quantum gravity
predicts.  In particular, we have derived two recursion relations
involving $N$ point and $N-1$ point correlation functions in a
topological quantum field theory with the semirigid geometry.  These
same relations partially characterize amplitudes of the one matrix
model at its topological critical point.
Thus, there is hope for one to continue to
show the rest
of the recursion relations involving Riemann surfaces with different
genera are reproduced in the topological semirigid gravity.  It is
only then that one can say topological semirigid gravity and the one
matrix model are equivalent.

Since intersection theory on moduli space and the one matrix model at
its topological point
are equivalent in the Kontsevich model and since these intersection
numbers satisfy the axioms extracted from topological quantum field
theories, a field theory of topological gravity
such as the semirigid formulation may be equivalent to the one matrix model.
The results of \dil\ and this paper give a concrete
example of how a field theory of
topological gravity and the one matrix model can be equivalent as
suggested by the intersection theory.

One can also imagine coupling topological matter to semirigid gravity
and computing
correlation functions including the matter fields.  We can then
compare to the higher matrix models or the one matrix model at a
different critical point and see if any of the recursion relations in
the matrix models are reproduced.  This will help us sort out what
these matrix models correspond to.

Finally, a comment is in order on a string theory
interpretation of this 2-d semirigid gravity.  Since the Liouville
mode decouples in the semirigid pure gravity, there is no constraint
analogous to the critical dimension $c=26$ on the type of topological
matter we may couple.
Hence, we may be able to construct a semirigid string
theory that lives in a four dimensional spacetime.

\vskip 0.5in

I would like to thank Gabriel Cardoso, Mark Doyle, Suresh
Govindarajan, Steve Griffies, and especially Philip
Nelson for valuable discussions.  I would also like to thank Philip
Nelson for making numerous suggestions to this manuscript.
This work was supported in part by DOE grant DOE-AC02-76-ERO-3071 and
NSF grant PHY88-57200.

\listrefs
\figures
\fig1{A coordinate patch $U$ with coordinates $(z,\theta)$ and origin
at $O$ on
the semirigid surface $\Sigmah$ is shown.  All odd coordinates are suppressed.
A local coordinate $\sigma=\zp$
centered at $P=(r,\rho)$ is given by $\sigma = f(z-r-\theta\rho)$ where $f$ is
some holomorphic function in the coordinates $(z,\theta)$.
A family of local coordinates at $P$ can be obtained if we now let $f$ be
parametrized by the moduli $({\vec m}, {\vec \zeta})$ of the
once-punctured surface at $P$. }

\fig2{A semirigid surface $\Sigmah$ with a unit disk $|\zp | <1$ removed
$(\Sigmah,\zp)$ is shown.  A change of the local coordinate $\zp$ to
$z'_P(\cdot)=(1+v)\zp$ is performed and then the boundaries of
$(\Sigmah, z'_P(\cdot))$ and the unit disk $D$ are identified.  A state
$|\Psi\>$ is inserted at the puncture $P$ on the disk.}

\bye